\documentclass[preprintnumbers,floatfix,superscriptaddress,pra,twocolumn,showpacs]{revtex4-1}
\usepackage{amsmath,amsthm,amssymb}
\usepackage{graphicx}
\graphicspath{ {images/} }
\usepackage{dcolumn}
\usepackage{bbold}
\usepackage{hyperref} 
\usepackage{braket}
\usepackage{xcolor}
\usepackage{dsfont}
\usepackage{setspace}
\usepackage{multirow}
\usepackage{booktabs}
\usepackage[font=footnotesize,justification=raggedright]{caption}
\usepackage[normalem]{ulem}

\newcommand{\figref}[1]{Fig.~\ref{#1}}

\newcommand{\secref}[1]{Sec.~\ref{#1}}
\newcommand{\appref}[1]{Appendix~\ref{#1}}

\DeclareGraphicsExtensions{.pdf,.png}
\newcommand{\ketbra}[3][]{\mathinner{\lvert#2\rangle\langle #3\rvert}_{#1}}
\newcommand{\proj}[2][]{\ketbra[#1]{#2}{#2}}
\theoremstyle{theorem}
\newtheorem{theorem}{Theorem}
\newtheorem*{theorem*}{Theorem}
\newtheorem{property}{Property}
\newcolumntype{C}[1]{>{\centering\arraybackslash}m{#1}}
\DeclareMathOperator{\Tr}{Tr}

\begin{document}

\title{Towards an equivalence between maximal entanglement\\and maximal quantum nonlocality}

\author{Victoria Lipinska}\email{v.lipinska@tudelft.nl}\affiliation{QuTech, Delft University of Technology, Lorentzweg 1, 2628 CJ Delft, The Netherlands}\affiliation{ICFO-Institut de Ciencies Fotoniques, The Barcelona Institute of Science and Technology, 08860 Castelldefels (Barcelona), Spain}
\author{Florian J. Curchod}\email{Florian.Curchod@icfo.eu}\affiliation{ICFO-Institut de Ciencies Fotoniques, The Barcelona Institute of Science and Technology, 08860 Castelldefels (Barcelona), Spain}
\author{Alejandro M\'attar}\affiliation{ICFO-Institut de Ciencies Fotoniques, The Barcelona Institute of Science and Technology, 08860 Castelldefels (Barcelona), Spain}
\author{Antonio Ac\'in}\affiliation{ICFO-Institut de Ciencies Fotoniques, The Barcelona Institute of Science and Technology, 08860 Castelldefels (Barcelona), Spain}
\affiliation{ICREA-Instituci\'o Catalana de Recerca i Estudis Avan\c cats, Lluis Companys 23, 08010 Barcelona, Spain}

\date{\today}

\begin{abstract}
While all bipartite pure entangled states are known to generate correlations violating a Bell inequality, and are therefore nonlocal, the quantitative relation between pure-state entanglement and  nonlocality is poorly understood. In fact, some Bell inequalities are maximally violated by non-maximally entangled states and this phenomenon is also observed for other operational measures of nonlocality. In this work, we  study a recently proposed measure of nonlocality defined as the probability that a pure state displays nonlocal correlations when subjected  to random measurements. We first prove that this measure satisfies some natural properties for an operational measure of nonlocality. Then, we show that for pure states of two qubits the measure is monotonic with entanglement for all correlation two-outcome Bell inequalities: for all these inequalities, the more the state is entangled, the larger the probability to violate them when random measurements are performed. Finally, we extend our results to the multipartite setting.
\end{abstract}

\maketitle

\section{\label{sec:intro}Introduction}

Entanglement, one of the key features of quantum theory, is an intrinsic property of the states describing joint quantum systems. Performing local measurements on the parts of entangled systems enables two distant parties to generate correlations that are in contradiction with the assumption of local realism~\cite{Bell1964a}. The resulting measurement statistics is commonly referred to as nonlocal. The presence of nonlocality is usually witnessed through the violation of a Bell inequality~\cite{Bell1964a,BellReview}, which, in turn, certifies the presence of entanglement in the underlying quantum system without any further assumptions or modeling of the experimental setup. However, understanding the exact relation between entanglement and nonlocality is not straightforward. It is unclear, for example, whether ``more" entanglement leads to ``more" nonlocality and, related to this, what is a good quantifier of nonlocality. In this work, we tackle this problem and analyze the connection between entanglement and nonlocality under a recently developed measure of nonlocality given by the probability that random measurements performed on a given state $\ket{\psi}$ generate nonlocal statistics.

Over the years, explaining the relation between entanglement and nonlocality has been the focus of attention for many works. Werner first revealed the subtlety of the problem by explicitly constructing a family of mixed entangled states that cannot violate any Bell inequality when subjected to projective measurements~\cite{Werner89}. Werner's result was later extended to general measurements in~\cite{Barrett}. For pure states, the situation seemed to clarify since Gisin recognized that all pure entangled systems of any dimension display nonlocality when applying appropriate measurements to them~\cite{GISIN1991201}. All these results, initially derived for bipartite systems, were also generalised to the multipartite case~\cite{Popescu,toth,Augusiak2015,Bowles}.

At the quantitative level, the relation between entanglement and nonlocality is not fully understood even for bipartite pure states. Early work by Tsirelson demonstrated that the maximal quantum violation of the Clauser-Horne-Shimony-Holt (CHSH) inequality~\cite{CHSH1969}, the simplest Bell inequality, can only be achieved when making measurements on a two-qubit maximally entangled state~\cite{Cirel'son1980}. It was then natural to expect maximal entanglement to be indispensable to retrieve the maximal quantum violation of Bell inequalities. However, subsequent examples showed this intuition to be wrong: the maximal quantum violation of certain Bell inequalities crucially requires partial entanglement~\cite{Acin_NLinQutrits2002}, even when considering states of arbitrary Hilbert space dimension~\cite{Liang2011,Vidick2011}. Furthermore, the phenomenon of obtaining more nonlocality from less entanglement for pure states happened to occur not only for the amount of violation of a given Bell inequality. It was also observed for other measures of nonlocality, such as the robustness of nonlocality to noise~\cite{Acin_NLinQutrits2002}, losses~\cite{Eberhard1993}, statistical strength of Bell tests~\cite{Acin2005} and the simulation of quantum correlations with nonlocal resources~\cite{Brunner2005}. This apparent inequivalence of pure-state entanglement and nonlocality was dubbed \textit{anomaly} in~\cite{Methot} and this is the terminology adopted here.

Even if there is no fundamental requirement for maximal entanglement and maximal nonlocality to be in one to one correspondence, it is desirable to understand if these anomalies appear only as artefacts of the measure that is used. In that sense, it would be interesting to come up with an operational measure of quantum nonlocality that would be maximized by maximally entangled states. A step in this direction was made in~\cite{Brasil2015}, where the authors gave numerical results suggesting that the anomaly originally observed in \cite{Acin_NLinQutrits2002} with two-qutrit states violating maximally the Collins-Gisin-Linden-Massar-Popescu (CGLMP) Bell inequality~\cite{CGLMP2002} disappears when considering a novel measure of nonlocality. For a given (pure) quantum state $\ket{\psi}$, the value of the measure is the probability of violating a specific Bell inequality when random projective measurements are performed on the state. A state $\ket{\psi_1}$ is more nonlocal than a state $\ket{\psi_2}$, in the sense of the measure studied in~\cite{Brasil2015}, if by making random measurements on $\ket{\psi_1}$ there is a higher chance of generating nonlocal correlations than on $\ket{\psi_2}$. This is the type of measure of nonlocality that we study here, which we name \emph{nonlocal volume}.  More specifically, the authors of \cite{Brasil2015} numerically showed that the probability of violating the three-outcome CGLMP inequality with random projective measurements is maximal among all pure two-qutrit states when using a maximally entangled state. Thus, the new quantifier removes the original anomaly between entanglement and nonlocality identified in~\cite{Acin_NLinQutrits2002} for the CGLMP inequality with three outputs. Note that the probability of finding nonlocal correlations for qubit states was initially considered in \cite{Liang}.

While the above study offers a promising insight into a potential measure of nonlocality for which the original anomaly disappears, several crucial aspects were not addressed there. The main limitation of this measure is that a single Bell inequality is used to witness nonlocality in the correlations. But apart from the simplest Bell-CHSH case, in any Bell scenario there are many inequivalent families of Bell inequalities. It is, then, unclear why a single inequality should be tested and preferred over the rest.
In the context of the nonlocality measure, this limitation was lifted later when the authors of \cite{Gdansk} extended the numerical search of~\cite{Brasil2015} without assuming any \textit{a priori} fixed Bell inequality. Instead, they considered all the possible Bell inequalities in a given Bell scenario. Note that this approach is equivalent to checking whether the given correlations are nonlocal independently of a specific Bell inequality, which provides a much more operational result. They then performed an intense numerical exploration of many different Bell setups, seeing that in all of them the largest value of the nonlocal volume was obtained for the maximally entangled state.

All this numerical evidence suggests that the nonlocal volume, that is, the probability of generating nonlocal correlations when performing random local measurements on a quantum state, is a good candidate for a measure of nonlocality without anomalies. On the other hand, to our knowledge almost no analytical results are known using this new measure. The only analytical results we are aware of concern the simplest scenario for quantum nonlocality with its unique CHSH inequality~\cite{Liang,Gdansk}, where it is known that the nonlocal volume is a monotone of entanglement: the more entangled the state, the bigger its probability to violate a CHSH inequality with random measurements made on it. The nonlocal volume for the maximally entangled state of two qubits only was computed analytically in \cite{Liang} and was found to be $2(\pi-3) \approx 28.32\%$. The reason why so little is known so far about the nonlocal volume is that it is hard to deal with it in analytically as one typically needs to solve complicated integrals.

In our work, we give the first analytical results connecting maximal entanglement and nonlocality in terms of the nonlocal volume. We start by defining properly the measure and proving that it indeed has many of the desirable properties as measure of nonlocality for quantum states. Specifically, we show that it is invariant under local unitaries (LU) applied by each party on the state, that its value is strictly positive for all pure bipartite entangled states and that its value tends to one in the limit of infinite measurement settings, as expected. 

We then prove that no anomaly can occur for two-qubit states when considering scenarios based on \textit{{correlation} inequalities} (or XOR games { \cite{Cleve2004,Briet2013} }) involving any number of projective two-outcome measurements per site. More generally, we show that these particular inequalities are monotonic with the amount of entanglement in two-qubit pure states: the more the entanglement in the state, the larger its probability of violating these Bell inequalities when random measurements are made on it. This implies, in particular, that the maximally entangled state is always the most nonlocal according to this measure in these scenarios. We show that our results extend to the multipartite scenario for the Greenberger-Horne-Zeilinger (GHZ) family of states $\cos(\theta)\ket{0...0}+\sin(\theta)\ket{1...1}$.
Finally, we demonstrate by providing explicit examples that our proof technique cannot be extended to scenarios involving two-output Bell inequalities with marginal terms.

\section{The nonlocal volume}\label{subsec:nonlocality}

In the standard bipartite Bell scenario \cite{Bell1964a,BellReview}, two parties Alice (A) and Bob (B) share entangled systems and each performs local measurements on its shares in separate laboratories. A (respectively B) performs one out of $m_A$ ($m_B$) possible measurements on her (his) system, obtaining one out of $o_A$ ($o_B$) possible outcomes. The measurement choices of Alice and Bob are labeled by $x=1,\ldots,m_A$ and $y=1,\ldots,m_B$ and  the corresponding outcome by $a=1,\ldots,o_A$ and $b=1,\ldots,o_B$. The measurements each party performs are described by a set of orthogonal projectors $\{M_{a|x}\}$ and $\{N_{b|y}\}$ that sum up to the identity $\sum\limits_{a}M_{a|x} = \sum\limits_{b}N_{b|y} = \mathbb{1}$. 
The corresponding Bell test is then fully described by the set of joint conditional probability distributions $\mathbf{p} = \{p(ab|xy) \}$, also called correlations or behavior. {They are given by,
\begin{equation}\label{QuantumP}
p(ab|xy) = \Tr(M_{a|x} \otimes N_{b|y} \rho).
\end{equation}
}
The set of all correlations of the form \eqref{QuantumP} forms the set $\cal{Q}$ of quantum correlations.

Within the set of quantum correlations, one can identify the set of local correlations $\cal{L}$, which can be generated when the parties have access to shared randomness only, or local hidden variables~\cite{BellReview}, admitting a decomposition of the form: 
\begin{equation}
\label{LocDecomp}
p(ab|xy) = \sum_\lambda q_\lambda p_\lambda (a|x) p_\lambda(b|y).
\end{equation}
where $\sum_\lambda q_\lambda = 1$, $q_\lambda \geq 0$.
Correlations that do not admit such a decomposition are referred to as nonlocal and are usually witnessed through the violation of a Bell inequality, that is a linear function of the probabilities $\hat{I}\left( \mathbf{p} \right) \equiv \sum_{abxy} g_{ab}^{xy}p(ab|xy)$, where $g^{ab}_{xy}$ are real coefficients. The maximum of $\hat{I}$ over local correlations $\mathbf{p} \in \cal{L}$ \eqref{LocDecomp} is the local bound and denoted by $g_{\text{loc}}$ so the Bell inequality reads $\hat{I}\left( \mathbf{p} \right) \leq g_{\text{loc}}$. For some choice of coefficients $g^{ab}_{xy}$, there exist quantum correlations $\mathbf{p} \in \cal{Q}$ \eqref{QuantumP} violating the corresponding inequality $\hat{I}\left( \mathbf{p} \right) > g_{\text{loc}}$. The violation of a Bell inequality prevails today as the most explored measure of nonlocality and was found to have many applications within the scope of quantum information science \cite{acin2006b,EkertQKD,MayersSelfTesting,PironioRandomness,BellReview}.

Now, consider a quantum system shared by A and B in a pure state of two qubits written in its Schmidt basis:
\begin{equation}
\label{PureQubits}
 \ket{\psi_{\theta}} =  \cos\left(\theta\right) \ket{00} + \sin\left(\theta\right) \ket{11}
\end{equation}
parametrized by the angle $\theta \in [0,\frac{\pi}{4}]$. Gisin showed that one can find local measurements on any state of the form \eqref{PureQubits} with $\theta>0$ such that the generated correlations are nonlocal~\cite{GISIN1991201}.
A natural question is then: which one among all the states $\ket{\psi_{\theta}}$ is the \textit{most} nonlocal, in the sense of giving the largest Bell inequality violation? The question is troublesome as the answer typically depends on the scenario and on the Bell inequality considered.  The situation simplifies in the setup with two dichotomic-outcome measurements per side, where the violation of the CHSH inequality alone is both necessary and sufficient to witness nonlocality. The state maximally violating the CHSH inequality upon optimization over the measurements is the maximally entangled state $\ket{\phi^+}$ ($\theta=\pi/4$ in \eqref{PureQubits})~\cite{Cirel'son1980}. In fact, in this case there even exists a monotonous relation between entanglement and nonlocality~\cite{HorodeckiCHSH}: the more entangled the state is, the more it violates the CHSH inequality. 

Intuitively, one could expect a similar monotonous relation between entanglement and nonlocality to hold for inequalities in broader scenarios, for Bell tests involving more measurement choices and/or outcomes, or even in full generality. In~\cite{Acin_NLinQutrits2002}, however, it was found that the CGLMP inequality \cite{CGLMP2002} with $o_A=o_B=3$ outcomes and with a two-qutrit state of the form $\ket{\psi^\gamma_3} = \tfrac{1}{\sqrt{2+\gamma^2}} (\ket{00} + \gamma\ket{11} + \ket{22})$ with $\gamma \simeq 0.79$ achieves a higher violation than obtained with the two-qutrit maximally entangled state $\ket{\phi^+_3} = \tfrac{1}{\sqrt{3}} (\ket{00} + \ket{11} + \ket{22})$. Furthermore, this anomaly of obtaining more nonlocality from less entanglement happened to occur for states of arbitrary dimension~\citep{ZohrenGill}, and for other measures of nonlocality as well~\cite{Eberhard1993,Acin2005,Brunner2005}. Note that most of the previous results were not rigorous proofs of the existence of an anomaly, as they mostly consisted of numerical searches. But subsequent works, such as~\cite{Liang2011,NPA,Vidick2011}, proved some of these results analytically.

As mentioned, to fix the original anomaly detected in~\cite{Acin_NLinQutrits2002}, the authors of \cite{Brasil2015} 
considered a measure of nonlocality defined by the probability that the correlations generated from randomly chosen measurements made on a given state $\ket{\psi}$ violate any Bell inequality by any extent. More formally, one defines the set of variables $\Omega$ parametrizing all the measurements that two parties may perform. For instance, a two-outcome projective measurement $M_{a|x} \equiv M_{a|x}(\omega_1,\omega_2)$ can be parametrized by two angles $\omega_1,\omega_2$ in the Bloch sphere. For all the measurement parameters in $\Omega$ one then needs to check whether the generated behavior from the state $\ket{\psi}$ is nonlocal. The parameters that do lead to measurements giving nonlocal correlations when made on $\ket{\psi}$ can be arranged in the set $\mathcal{V}(\ket{\psi})$. We are interested in calculating the relative volume of the set $\mathcal{V}(\ket{\psi})$ with respect to the volume of the whole set $\Omega$. The reason for it is that it can be directly interpreted as the probability of obtaining nonlocal correlations with random measurements, i.e. $P_{NL}(\ket{\psi}) = \frac{\textnormal{vol}(\mathcal{V}(\ket{\psi}))}{\textnormal{vol}(\Omega)}$. {Note that the exact value of this probability depends on the value of the volumes, which, in turn, is a function of the measure chosen to sample the measurements. As discussed below, for projective measurements the sampling is naturally defined by the Haar measure, which is the only measure invariant under unitary operations. Moreover, we remark that some of our results are valid for any choice of measure.}

Equivalently, the nonlocal volume can be obtained by considering the following quantity
\begin{equation}\label{DefAnomaly}
P_{NL}(\ket{\psi}) = \int \textnormal{d}\Omega f(\ket{\psi},\Omega),
\end{equation}
where we integrate over the measurement parameters $\Omega$ according to the Haar measure. The function $f(\ket{\psi},\Omega)$ is an indicator function that takes the value 1 whenever the generated behavior is nonlocal and 0 otherwise:
\begin{equation}
\label{IndicFunction}
f(\ket{\psi},\Omega) = \begin{cases}
1 \hspace{0.3cm} \textrm{if $p(ab|xy)$ is nonlocal} \\
0 \hfill \textrm{otherwise}
\end{cases}
\end{equation}
Using this definition, the potential nonlocality of the generated behaviors $p(ab|xy) = \Tr(\ket{\psi}\bra{\psi} M_{a|x} \otimes M_{b|y})$ \eqref{IndicFunction} can be understood as witnessed by \textit{all possible} Bell inequalities for a given scenario. Note that this is equivalent to checking whether some given correlations admit a local decomposition \eqref{LocDecomp}. In that sense, the violation of a Bell inequality should be understood as a witness of nonlocality only, and not as a quantifier. Seen as witnesses, it is then important to consider the full set of possible inequalities in a setup, as it would otherwise be possible for nonlocal correlations to go undetected and lead to an underestimation of the nonlocal volume.

In general, explicitly evaluating the integral in \eqref{DefAnomaly} can be highly demanding. So far, analytical results exist only in the simplest bipartite case and for the CHSH inequality~\cite{Liang,Gdansk}. 
Nonetheless, the numerical results of \cite{Brasil2015,Gdansk} strongly suggest that the above measure may be able to remove the anomaly between nonlocality and entanglement. Indeed, extensive numerical computations show that the maximally entangled state is the one achieving the highest probability of obtaining nonlocal correlations with random measurements in all the explored cases.

\section{Properties of the measure}\label{sec:properties}

The nonlocal volume \eqref{DefAnomaly} aims at measuring how nonlocal pure states are in order to compare them. As such, we clearly want this measure to fulfill a basic set of conditions to consider it an operational measure of nonlocality. In this section we list some of the desired properties and formally prove that the nonlocal volume satisfies them.

\begin{property}\label{LUInvariance}
The nonlocal volume \eqref{DefAnomaly} is invariant under local unitaries applied on the state if one uses the Haar measure for the integration:
\begin{equation}
P_{NL}(V_1 \otimes V_2 ~\rho~ V_1^\dagger \otimes V_2^\dagger) = P_{NL}(\rho) \hspace{1cm} \forall \hspace{0.2cm} V_1, V_2
\end{equation}
where $V_1,V_2$ are local unitary transformations applied by the parties to their share of the state. 
\end{property}

\begin{proof}
\begin{equation}\begin{split}\label{proof1}
P_{NL}(V_1 \otimes V_2 ~\rho~ V_1^\dagger \otimes V_2^\dagger)\\
= \int \textnormal{d}\Omega f(V_1 \otimes V_2 ~\rho~ V_1^\dagger \otimes V_2^\dagger,\Omega)
\end{split}\end{equation}
Now, using the cyclicity of the trace operator:
\begin{equation}\begin{split}
\Tr(V_1 \otimes V_2 ~\rho~ V_1^\dagger \otimes V_2^\dagger M_{a|x}^{\Omega}\otimes N_{b|y}^{\Omega})\\ = \Tr(\rho \hspace{0.1cm} V_1^\dagger  M_{a|x}^{\Omega}V_1 \otimes V_2^\dagger N_{b|y}^{\Omega} V_2)
\end{split}
\end{equation}

Finally, making the substitution $\Omega \rightarrow \Omega '$ such that $M_{a|x}^{\Omega '} \otimes N_{b|y}^{\Omega '} =  V_1^\dagger  M_{a|x}^{\Omega}V_1 \otimes V_2^\dagger N_{b|y}^{\Omega} V_2$ and the fact that, if using the Haar measure, $\textnormal{d}\Omega ' = \textnormal{d}\Omega$ (the elements of integration are invariant under LU) leads to the desired result.
\end{proof}

\begin{property}\label{Positivity}
For all pure bipartite entangled states $\ket{\psi}_{\textrm{ent}}$ in a setup with at least two choices of two-outcome measurements, the nonlocal volume \eqref{DefAnomaly} is strictly positive:
\begin{equation}
P_{\textrm{NL}}(\ket{\psi}_{\textrm{ent}}) > 0
\end{equation}
and thus:
\begin{equation}
P_{\textrm{NL}}(\ket{\psi}) = 0
\end{equation}
if and only if the state $\ket{\psi}$ is separable.
\end{property}
\begin{proof}
To see this, first consider the space $\Omega$ of parameters parameterizing all the local measurements. For example, a two-outcome projective measurement on a qubit state can be parameterized by two angles $\omega_1,\omega_2$ in the Bloch sphere. From \cite{GISIN1991201}, we know that for any pure entangled state $\ket{\psi}_{\textrm{ent}}$ (of any dimension) there exist certain values of the parameters such that the measurements performed on the state generate correlations that are nonlocal in the simplest setup with $x,y,a,b = 1,2$, i.e. $\cal{V}(\ket{\psi}_{\textrm{ent}}) \neq \emptyset$. We still need to show that the set of parameters leading to nonlocal correlations $\cal{V}(\ket{\psi}_{\textrm{ent}})$ is not of volume zero. Note that since the local correlations form a closed set, for any fixed state the set of measurement parameters leading to local correlations \eqref{LocDecomp} is also closed. This implies that the (disjoint) sets of parameters leading to nonlocal correlations are open. In particular there is always a ball around any nonlocal point in this space of parameters that contains parameters leading to nonlocal correlations as well. For any fixed pure entangled state is then clear that starting from any nonlocal quantum correlations one can slightly perturb all the parameters $\omega$ and still generate nonlocal correlations.
\end{proof}

\begin{property}\label{Assymptotic}
For any pure bipartite entangled state $\ket{\psi}_{\textrm{ent}}$, the nonlocal volume \eqref{DefAnomaly} tends to unity when the number of measurement choices tends to infinity:
\begin{equation}
P_{NL}(\ket{\psi}_{\textrm{ent}}) \xrightarrow[m_B \rightarrow \infty]{m_A \rightarrow \infty} 1
\end{equation}
\end{property}
\begin{proof}
From property \eqref{Positivity}, we know that for the pure state $\ket{\psi}_{\textrm{ent}}$ and in the setup with $x,y,a,b = 1,2$ the nonlocal volume is strictly larger than zero $P_{NL}(\ket{\psi}_{\textrm{ent}}) = \epsilon > 0$. The probability that the generated correlations $\{p(ab|xy)\}_{x,y = 1,2}$ are local for random measurements is then $P_{\textrm{loc}} = 1-P_{NL} = 1-\epsilon$. Now, with additional measurement settings, say $x = 3,4$ and $y = 3,4$, the correlations $\{p(ab|xy)\}_{x,y = 3,4}$ also has a probability $P_{\textrm{loc}} = 1-\epsilon$ of being local, independently of $\{p(ab|xy)\}_{x,y = 1,2}$. By repeating the argument and thus increasing the number of measurements choices, the probability that \textit{all} two-settings correlations $\{p(ab|xy)\}_{x,y = 2k-1,2k}$ with $k=1,2,...$ are local is:
\begin{equation}
P^k_{\textrm{loc}}(\ket{\psi}_{\textrm{ent}}) = (1-\epsilon)^k
\end{equation}
Remark that if any of these two-settings correlations are nonlocal, then clearly the full correlations also are nonlocal. This implies that
\begin{equation}
P_{NL}(\ket{\psi}_{\textrm{ent}}) \geq 1- P^k_{\textrm{loc}}(\ket{\psi}_{\textrm{ent}}) = 1 - (1-\epsilon)^k  \xrightarrow[]{k \rightarrow \infty} 1,
\end{equation}
{which means that the lower bound on $P_{NL}$ goes to 1 as $k\rightarrow \infty$. Moreover, we have that $k \rightarrow \infty $ implies $m_A,m_B \rightarrow \infty$, which yields the desired result. }
\end{proof}

Note that the numerical evidence suggesting Property \ref{Assymptotic} of the nonlocal volume had been found in \cite{Gdansk,Shadbolt2011}.

{Finally, let us comment on the generalisation of properties \ref{Positivity} and \ref{Assymptotic} to bipartite mixed entangled states that are nonlocal, i.e. mixed states for which one can find local measurements such that the generated correlations violate a Bell inequality. Clearly, if the mentioned measurements can be found in a scenario involving finite numbers of measurements settings $m_A,m_B$, then one can obtain properties similar to \ref{Positivity} and \ref{Assymptotic} for a given mixed nonlocal state $\rho$. In a scenario with at least $m_A$ and $m_B$ settings -- instead of $m_A,m_B = 2$ for a pure entangled state, $P_{NL}(\rho) > 0$ since one can always slightly perturb the mentioned measurements and still generate nonlocal correlations. This observation comes from the fact that the set of local correlations is also closed in that scenario. Property \ref{Assymptotic} also holds for any mixed nonlocal state and only the proof needs to be adapted. One now considers $k$ disjoints sets consisting of $m_A,m_B$ measurements each (instead of $m_A,m_B=2$ for pure entangled states) and by taking $k$ large enough the probability that \textit{all} the correlations $\{p(ab|xy)\}^{x=(k-1) m_A,(k-1) m_A+1,...,k m_A -1}_{y=(k-1) m_B,(k-1) m_B+1,...,k m_B -1}$ for k = 1,2,... are local tends to zero, implying that the probability of the full correlations being nonlocal tends to one with growing $k$.}

\section{The nonlocal volume using correlation Bell inequalities is a monotone of entanglement}\label{subsec:two_corr+thrm}

Having proven some of the properties of the measure \eqref{DefAnomaly}, we proceed to analyzing the nonlocal volume of different entangled states. We are still unable to compute $P_{NL}(\ket{\psi})$ explicitly from Definition \ref{DefAnomaly}. Therefore, we approach the problem alternatively and study whether there exist inclusion relations among the sets $\mathcal{V}(\ket{\psi})$ of measurements leading to nonlocal correlations when made on different states. Indeed, if the set of measurements $\mathcal{V}(\ket{\psi_1})$ leading to nonlocal correlations on the state $\ket{\psi_1}$ is \textit{included} in the set $\mathcal{V}(\ket{\psi_2})$ for the state $\ket{\psi_2}$, $\mathcal{V}(\ket{\psi_1}) \subseteq \mathcal{V}(\ket{\psi_2})$, then obviously $P_{NL}(\ket{\psi_1}) \leq P_{NL}(\ket{\psi_2})$.
Crucially, we show that in many situations, namely when witnessing nonlocality with correlation (see below \eqref{FullCorrel}) inequalities only, the set of measurements $\mathcal{V}(\ket{\psi_{\theta_1}})$ leading to nonlocal correlations on a pure two-qubit entangled state $\ket{\psi_{\theta_1}}$ is included in the set $\mathcal{V}(\ket{\psi_{\theta_2}})$ if $\ket{\psi_{\theta_1}}$ is less entangled than $\ket{\psi_{\theta_2}}$. We thus prove that the nonlocal volume of {correlation} Bell inequalities is a monotone of entanglement in the case of qubit states and two-outcome projective measurements. 

We work in Bell scenarios with two-outcome measurements and any number of measurement settings per party. Labeling the measurements outcomes $a,b = \pm 1$, the correlations in this scenario can be parametrized as
\begin{equation}
 p(ab|xy)=\frac{1}{4}\left(1+a\langle A_x \rangle + b\langle B_y \rangle+ab\langle A_x B_y \rangle\right) ,
 \end{equation} 
where $\langle A_x \rangle = \sum\limits_{a = \pm 1} a~p_A(a|x)$ are Alice's local expectation value depending on her marginal distribution $p_A(a|x)=\sum\limits_b p(ab|xy)$, and similarly for Bob's $\langle B_y \rangle$. The terms $\langle A_x B_y \rangle=\sum\limits_{a,b=\pm 1}ab~p(a b|xy)$ are known as two-body correlators. In this scenario, {correlation} or {full-correlator} Bell inequalities (or even \textit{XOR} games) for two outcomes are those in which only these last terms appear and hence can be written as 
\begin{equation}
\label{FullCorrel}
\hat{I}^{\langle .. \rangle} = \sum\limits_{xy} g_{xy} \langle A_x B_y \rangle \leq g_\textnormal{loc}
\end{equation}
where $g_\textnormal{loc}$ is the local bound \eqref{LocDecomp}.

For any {correlation} Bell inequality $I^{\langle .. \rangle}$ and for local measurements $M_{a|x}$ and $N_{b|y}$, one can define the associated Bell operator (acting at the level of the states):
\begin{equation}
\label{OPB}
B_{I^{\langle .. \rangle}} = \sum\limits_{xy} g_{xy} A_{x} \otimes B_{y},
\end{equation}
where we defined the observables $A_x=M_{+1|x}-M_{-1|x}$, $B_y=N_{ +1|y}-N_{-1|y}$. For a given state $\rho$, the value of the Bell inequality then reads 
\begin{equation}
\label{BellOP}
I^{\langle .. \rangle}(\rho) = \Tr (\rho B_{I^{\langle .. \rangle}}) .
\end{equation}

Next, we present our main result under the form of a theorem. Our result holds for any number of 2-outcome projective measurements performed by Alice and Bob.

\begin{theorem}\label{THRM:Bell}
Consider any {correlation} Bell inequality $\hat{I}^{\langle .. \rangle} = \sum\limits_{xy} g_{xy} \langle A_x B_y \rangle \leq g_\textnormal{loc}$ \eqref{FullCorrel} with $g_\textnormal{loc}$ being the local bound. $A$ and $B$ measure the local observables $\{A_{x}\}$ and $\{B_{y}\}$ respectively, defining the associated Bell operator $B_I^{\langle .. \rangle} = \sum_{xy} g_{xy} A_{x} \otimes B_{y}$  \eqref{BellOP}. Consider two pure two qubit states $\ket{\psi_{\theta_1}}$ and $\ket{\psi_{\theta_2}}$ with $\theta_1,\theta_2 \in [0,\frac{\pi}{4}]$ \eqref{PureQubits} and $\theta_2 > \theta_1$ such that $\ket{\psi_{\theta_1}}$ violates the inequality, that is $\Tr (\proj{\Psi_{\theta_1}} \hat{B}_{I^{\langle .. \rangle}}) > g_\textnormal{loc}$. Then:

\begin{equation}\label{MainResult_ghz}
\Tr (\proj{\Psi_{\theta_2}} \hat{B}_{I^{\langle .. \rangle}}) > \Tr (\proj{\Psi_{\theta_1}} \hat{B}_{I^{\langle .. \rangle}}).
\end{equation}
\end{theorem}
In words, if a {correlation} Bell inequality $I^{\langle .. \rangle}$ is violated by correlations generated when $A$ and $B$ measure the local observables $\{A_{x}\}$ and $\{B_{y}\}$ respectively on a pure partially entangled two qubit state $\ket{\psi_{\theta_1}}$, then the same inequality with the same measurements gives a strictly larger violation when acting on any other pure entangled two qubit state $\ket{\psi_{\theta_2}}$ with more entanglement $\theta_2 > \theta_1$.

\begin{proof}
Observe that $\ket{\psi_{\theta}}$ can always be written as
\begin{equation}\label{DecOfTheta}
\ket{\psi_{\theta}} =
\left(
\tfrac{\cos \theta + \sin \theta}{\sqrt{2}} \mathbb{1} + \tfrac{\cos \theta - \sin \theta}{\sqrt{2}} \sigma_z
\right)
\otimes \mathbb{1}\ | \phi^+ \rangle.
\end{equation}
Denote by $\hat{B}_{I^{\langle\cdot \cdot \rangle}}$ the Bell operator associated to the inequality $I^{\langle .. \rangle}$ \eqref{OPB} for the given local measurements.
Since the inequality $I^{\langle .. \rangle}$ contains only full-body correlators, it does not involve marginal terms and thus the decomposition of the Bell operator $\hat{B}_{I^{\langle\cdot \cdot \rangle}}$ in the Pauli basis does not contain terms proportional to $\mathbb{1}\otimes \mathbb{1}$, $\mathbb{1}\otimes \sigma_{i}$ and $\sigma_{i}\otimes \mathbb{1}$, for $i=x,y,z$. Using this fact and expression \eqref{DecOfTheta}, the Bell violation for state \eqref{DecOfTheta}, $b_{\theta} \equiv \Tr(\proj{\psi_{\theta}} | \hat{B}_{I^{\langle\cdot \cdot \rangle}})$, reads
\begin{equation}\label{ConvexDecOfTheta}
b_{\theta} = \frac{b_+ + b_-}{2} + \frac{\sin 2\theta}{2} (b_+ - b_-) >g_\textnormal{loc},
\end{equation}
where $b_{\pm} \equiv \bra{\phi^{\pm}}\hat{B}_{I^{\langle\cdot \cdot \rangle}}  \ket{\phi^{\pm}}$ denotes the expectation value of $\hat{B}_{I^{\langle\cdot \cdot \rangle}}$ on the maximally entangled state $\ket{\phi^{\pm}}=\frac{1}{\sqrt{2}}(\ket{00}\pm\ket{11})$.

By hypothesis we have that when $\theta=\theta_1$
\begin{equation}\label{btheta}
\setstretch{1.5}
\begin{split}
b_{\theta_1} \equiv \Tr(| \psi_{\theta_1} \rangle \langle \psi_{\theta_1} | \hat{B}_{I^{\langle\cdot \cdot \rangle}})
 > g_\textnormal{loc},
\end{split}
\end{equation}
The term $\frac{b_+ + b_-}{2}$ can be understood (by linearity of the trace) as the expectation value of $ \hat{B}_{I^{\langle\cdot \cdot \rangle}}$ on the separable state 
\begin{equation}
\frac{1}{2}(\proj{00}+\proj{11})
\end{equation}
and is thus necessarily smaller or equal to $g_\textnormal{loc}$.
But since $\sin 2\theta$ is positive for $\theta \in [0,\frac{\pi}{4}]$, Eq. \eqref{ConvexDecOfTheta} necessarily implies that $b_+ > b_-$. Now, because of this property and the fact that $\sin 2\theta$ is monotonically increasing for $\theta \in [0,\frac{\pi}{4}]$, the proof of the theorem follows.
\end{proof}

Put differently, the theorem shows that when using {correlation} Bell inequalities $\hat{I}^{\langle .. \rangle}$ \eqref{FullCorrel} only to witness nonlocal correlations, the set of measurements $\cal{V}^{\langle\cdot \cdot \rangle}$$(\ket{\psi_{\theta_1}})$ generating nonlocal behaviors when performed on $\ket{\psi_{\theta_1}}$ is included in the set of measurements $\cal{V}^{\langle\cdot \cdot \rangle}$$(\ket{\psi_{\theta_2}})$ leading to nonlocal correlations when performed on any state $\ket{\psi_{\theta_2}}$ with more entanglement $\theta_2 > \theta_1$. This, in particular, implies that no anomaly can ever occur in these cases.

We now want to show that the inclusion relation $\cal{V}^{\langle\cdot \cdot \rangle}$$(\ket{\psi_{\theta_1}}) \subset \cal{V}^{\langle\cdot \cdot \rangle}$$(\ket{\psi_{\theta_2}})$ is strict. In the setup with two measurement choice with two outcomes, the violation of the CHSH inequality
\begin{equation}\begin{split}\label{CHSH}
\langle A_0B_0 \rangle + \langle A_0B_1 \rangle +\langle A_1B_0 \rangle - \langle A_1B_1 \rangle \leq 2
\end{split}\end{equation}
is both necessary and sufficient for witnessing nonlocality in the correlation. In that scenario one can check that $A$ and $B$ measuring the following observables:
\begin{equation}\begin{split}\label{MeasurementsInclusion}
A_{x=0} = \sigma_x \hspace{1cm} B_{y=0} = \cos(\xi)\sigma_x + \sin(\xi)\sigma_z\\
A_{x=1} = \sigma_z \hspace{1cm} B_{y=1} = \cos(\xi)\sigma_x - \sin(\xi)\sigma_z
\end{split}\end{equation}
with $\xi \in [0,\frac{\pi}{2}]$ on a pure two qubit state $\ket{\psi_{\theta}}$ \eqref{PureQubits} gives:
\begin{equation}\begin{split}\label{CHSHangles}
\textrm{CHSH}(\theta,\xi) = 2 \bigg( \sin(\xi)+\sin(2\theta)\cos(\xi) \bigg)
\end{split}\end{equation}
which for $\theta = \frac{\pi}{4}$ and all $\xi>0$ is larger than $2$ (the local bound). Now, for another value of $\theta$, the inequality is violated if
\begin{equation}\begin{split}\label{CHSHcond}
\sin(2\theta) > \frac{1-\sin(\xi)}{\cos(\xi)}
\end{split}\end{equation}
implying in particular that for any $\theta_2 > \theta_1$ -- i.e. $\sin(2\theta_2) > \sin(2\theta_1)$ -- one can find an angle $\bar{\xi}$ such that $\textrm{CHSH}(\theta_2,\bar{\xi}) > 2$ but $\textrm{CHSH}(\theta_1,\bar{\xi}) \leq 2$. In the end, this allows us to conclude that the inclusion of sets is strict
\begin{equation}\label{result1}
\mathcal{V}^{\langle\cdot \cdot \rangle}(\ket{\psi_{\theta_1}})  \subset \mathcal{V}^{\langle\cdot \cdot \rangle}(\ket{\psi_{\theta_2}}) ,
\end{equation}
Consequently, and in the spirit of definition \eqref{DefAnomaly}, it follows that:
\begin{equation}\label{result2}
P_{\textrm{NL}}^{\langle\cdot \cdot \rangle}(\ket{\psi_{\theta_1}}) \leq P_{\textrm{NL}}^{\langle\cdot \cdot \rangle}(\ket{\psi_{\theta_2}})  \leq P_{\textrm{NL}}^{\langle\cdot \cdot \rangle}(\ket{\phi^+}),
\end{equation}
where $P_{\textrm{NL}}^{\langle\cdot \cdot \rangle}(\ket{\psi_{\theta_1}})$ is defined in the same fashion as in \eqref{DefAnomaly}, but assuming that nonlocal correlations may only be witnessed by {correlation} inequalities. Crucially, and in sound contrast with previous works~\cite{Liang2011,Brasil2015,Gdansk}, our results are valid for any number of measurement settings, and -- interestingly -- as well as for any measurement sampling in $\eqref{DefAnomaly}$ (not only for the Haar measure). It is also worth noting that in many scenarios facet inequalities --- those delimiting the local set $\cal{L}$ --- are {correlation} inequalities, meaning that our result applies to a very broad class of inequalities \cite{WernerWolf:22n} in any scenario \cite{allCHSH,WernerWolf,QuantMultiNL,EntDepth,Flo}.

Furthermore, our result enables us to draw conclusions beyond the fundamental study of the relation between entanglement and nonlocality. {In a situation where one wants to check whether given measurements are useful to violate a correlation Bell inequality with two-qubit states, a necessary and sufficient condition is that they generate nonlocal correlations when performed on the maximally entangled state. Indeed, if the measurements do not generate nonlocality with the maximally entangled state, they will not generate nonlocality with any other less entangled state. What is more, since the maximally entangled state is the one with the highest probability to reveal nonlocality (up to any extend), it is the best choice to succeed in any Bell test using {correlation} inequalities and two-qubit states with poor control over the measurement bases. This is of particular interest for experimental setups where aligning reference frames is troublesome.}

Before concluding, we would like to connect this result with previous works. Tsirelson showed that the maximal violation of a two-outcome correlation Bell inequality is obtained for a maximally entangled state~\cite{Tsirelson}. However, this state is not necessarily of two qubits. In fact, there are known examples of two-outcome correlation Bell inequalities whose maximal violation requires systems of dimension larger than 2~\cite{Vertesi}. When discussing qubits, the maximal violation of correlation Bell inequalities is obtained by a maximally entangled state, as the Bell operator is always diagonal in a given Bell basis. Recall, however, that this has a priori no implications for the nonlocal volume, as maximal violation and nonlocal volume are unrelated quantities. For example, the maximally entangled state does not give the maximal violation of the CGLMP inequality,  but it does maximise the nonlocal volume. Note, however, that our theorem goes beyond proving that the maximal qubit violation is obtained by a Bell state: for fixed Schmidt bases, which do not necessarily coincide with those of the maximally entangled state providing the maximal qubit violation, the largest violation is obtained by a maximally entangled state.

\section{Possible bipartite generalisations of the result}\label{sec:extensions}

Our next objective is to discuss possible extensions of our result. In \secref{subsec:two_corr+thrm} we made three important assumptions: \textit{i)} the Bell inequality is a {correlation} inequality, \textit{i.e.} without marginal terms (single-body correlators), \textit{ii)} only two-qubit pure states  were considered, and \textit{iii)} only extremal (thus projective) measurements were considered. 

As far as assumption \textit{i)} is concerned, a numerical search provided us with an analytical counter-example consisting of measurements generating correlations violating a Bell inequality when performed on $\ket{\psi_\theta}$ for $\theta=\frac{3\pi}{16}$ but generating local correlations when performed on $\ket{\phi^+}$ in the $[3,4,2,2]$ scenario. We verified that the violated Bell inequality indeed contains marginal terms $\langle A_x \rangle$ and $\langle B_y \rangle$ as expected. We refer the reader to \appref{APP:counterexample} for the exact construction. This counterexample closes the possibility to generalize our theorem onto general Bell inequalities including single-body correlators. Therefore, the sets $\mathcal{V}(\ket{\psi_\theta})$ and $\mathcal{V}(\ket{\phi_+})$ are not contained one into another and we can not conclude on the relation between $P_{NL}({\ket{\psi_{\theta}}})$ and $P_{NL}({\ket{\phi^+}})$ based on inclusion relations between these sets. 

\begin{figure}[!h]
\centering
\includegraphics[width=0.5\textwidth]{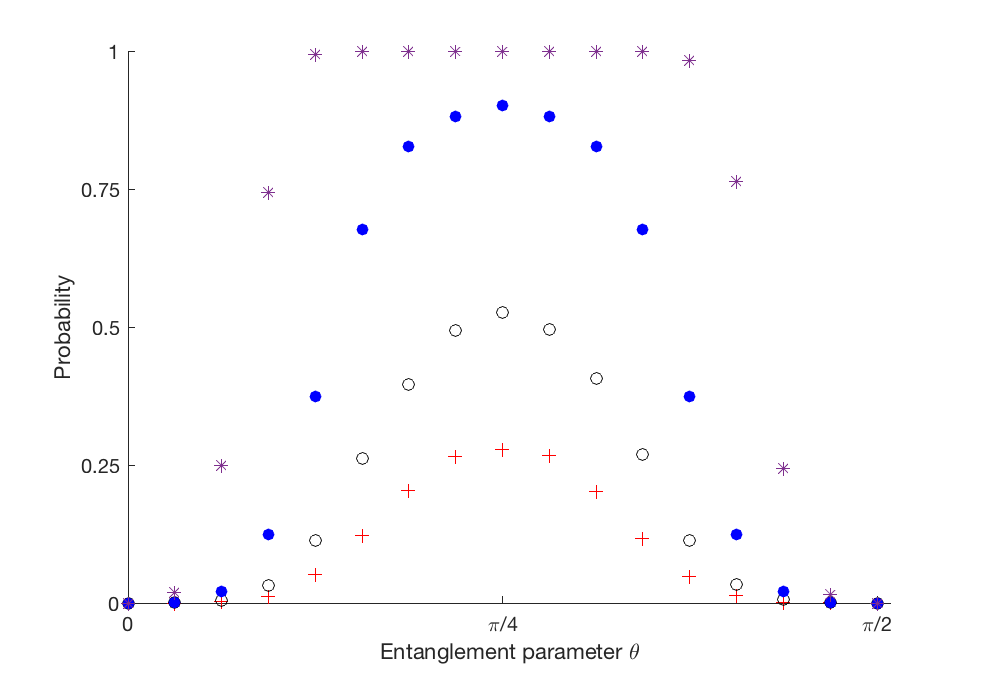}
\caption{Probability of obtaining nonlocal correlations with uniformly random measurements as a function of the entanglement parameter $\theta$. For clarity of the image the range of $\theta$ has been extended to $\tfrac{\pi}{2}$ due to symmetry of the state $\ket{\psi_\theta}$. Measurement scenarios: ($+$) -- [2,2,2,2], ($\circ$) -- [2,3,2,2], ($\bullet$) -- [3,4,2,2], ($\ast$) -- [8,8,2,2].}
\label{FIG:probability_vs_theta}
\end{figure}

As it is impossible to prove an analog of our main theorem for general two-outcome Bell inequalities including marginals, we numerically computed the value of the nonlocal volume \eqref{DefAnomaly} for arbitrary two-qubit states and different Bell scenarios. In \figref{FIG:probability_vs_theta}, we provide numerical evidence for a wide range of scenarios that indicate that the probability of generating nonlocal correlations from random measurements is always the largest when measuring the maximally entangled state. We conjecture that the relation $P_{NL}({\ket{\psi_{\theta}}}) \leq P_{NL}({\ket{\phi^+}})$ \eqref{DefAnomaly} holds in general. Note that similar numerical results were obtained in \cite{Gdansk}.

In order to relax assumption \textit{ii)} one can study states in systems of arbitrary dimension $\mathbb{C}^d \times \mathbb{C}^d$. Note that in these systems, the ordering induced between entangled states is partial at the single-copy level, as there are pairs of states that can not be deterministically transformed one into another in either way by local operations and classical communication (LOCC)~\cite{Nielsen}. So, it is unclear which entanglement quantifier would be a good candidate to be in correspondence with the nonlocal volume. The most natural candidate is the entanglement entropy, but it is a quantity that becomes especially relevant in the many-copy regime~\cite{ententropy}. Despite all these issues, there is a clear notion of maximally entangled state. Thus, the most natural working conjecture is that this state maximizes the nonlocal volume. Numerical searches already performed in \citep{Gdansk} indicate that this may be the case. More precisely, the authors considered states $\tfrac{1}{\sqrt{2+\gamma^2}}(\ket{00} + \gamma \ket{11} + \ket{22})$ with parameter $\gamma \in [0,1]$ and found that the highest probability of obtaining nonlocality with randomly sampled measurements occurs for $\gamma = 1$. It is also interesting to consider weaker variants of this conjecture that may be easier to attack. For instance there is a notion of correlation function and correlation Bell inequality for scenarios involving measurements of more than two outputs { \cite{CGLMP2002, Salavrakos2017}}. Understanding whether Theorem \ref{THRM:Bell} generalizes to this partial case deserves further investigation.

As for assumption \textit{iii)}, extending our study to general measurements beyond projective is also interesting. Note, however, that in this case, it is less clear what the natural way of sampling measurements should be.

\section{nonlocal volume in the multipartite scenario}\label{sec:multipartite}

So far our analysis has focused on bipartite settings. Extending the problem to the multipartite case is also interesting and first numerical steps in this direction were presented in~\cite{Liang,Gdansk}. Here we provide the first analytical results. Note that in the multipartite case there is no notion of maximally entangled state~\cite{MREGS}. So it is not clear which state should be the natural candidate to maximise the nonlocal volume and it could even happen that the optimal state varies with the number of parties. In the following however we show that in a restricted multipartite scenario, it is possible to generalize our main result and conclude about the monotonicity of the measure for specific families of states and {correlation} Bell inequalities. 

In a multipartite scenario, $n$ parties share an entangled system of many particles. Each party $A_i$, $i=1,...,n$, performs a local measurement on its share of the system with measurement choice labeled $x_i=1,...,m_{A_i}$ and (dichotomic) outcome $a_i = 0,1$. As before, the measurements each party performs are described by a set of orthogonal projectors $\{ M^{(i)}_{a_i|x_i}\}$, which generate joint conditional probabilities $\mathbf{p} = \{p(a_1 \ldots a_n | x_1 \ldots x_n) \}$. Then, $p(\vec{a}|\vec{x}) \equiv p(a_1 \ldots a_n | x_1 \ldots x_n) = \Tr(M^{(1)}_{a_1|x_1} \otimes \ldots \otimes M^{(n)}_{a_n|x_n} ~ \rho)$. As in the bipartite scenario, a Bell inequality is a linear combination of the probabilities $\hat{I}^{\langle n \rangle}\left( \mathbf{p} \right) \equiv \sum\limits_{a_1...a_nx_1...x_n} g_{a_1...a_n}^{x_1...x_n}p(\vec{a}|\vec{x})$ and corresponds to a Bell operator acting at the level of the states $\hat{B}_{I^{\langle n \rangle}} = \sum\limits_{a_1...a_nx_1...x_n} g_{a_1...a_n}^{x_1...x_n} M^{(1)}_{a_1|x_1} \otimes \ldots \otimes M^{(n)}_{a_n|x_n}$.

For two-outcome measurements only, we can define full-body correlators
\begin{align}
\langle A_{x_1} ... A_{x_n} \rangle = \sum\limits_{a_1...a_n = 0,1} (-1)^{\sum\limits_{i=1}^n a_i}  p(\vec{a}|\vec{x}),
\end{align}
and a {correlation} inequality ({inequality with $n$-body correlators})
\begin{align}
I^{\langle n\rangle} = \sum_{x_1...x_n} \tilde{g}_{x_1...x_n} \langle A_{x_1} ... A_{x_n} \rangle.
\end{align}

As we mentioned, it is much harder in the multipartite setting to order (pure) states in terms of how entangled they are. To avoid the problem, we focus on a natural generalization of the bipartite pure states $\ket{\phi_{\theta}}$ \eqref{PureQubits} 
\begin{equation}\label{GHZ}
\ket{\Psi_{\theta}^n} = \cos \theta \ket{0}^{\otimes n} + \sin \theta \ket{1}^{\otimes n},
\end{equation}
where $\theta$ is the entanglement parameter whose value runs again from 0 to $\pi/4$. The maximally entangled state of this family, with $\theta = \frac{\pi}{4}$, is the GHZ state $\ket{\textrm{GHZ}^n} \equiv \ket{\Psi_{\theta=\frac{\pi}{4}}^n}$ as any other state in the family can be deterministically reached from it by LOCC. 

We now generalize Theorem \ref{THRM:Bell} to the multipartite setup for an even number of parties, correlation Bell inequalities and pure states in the GHZ family~\eqref{GHZ}. \\

\begin{theorem}\label{THRM:Bell_ghz}
Consider a {correlation} Bell inequality $I^{\langle n\rangle} = \sum\limits_{x_1...x_n} \tilde{g}_{x_1...x_n} \langle A_{x_1} ... A_{x_n} \rangle \leq g_\textnormal{loc}$ with $g_\textnormal{loc}$ being the local bound. Assume that the number of parties $n$ is even. Each party measures, locally, the observable $\{A^{(i)}_{x_i} \equiv M^{(i)}_{a_i=0|x_i} - M^{(i)}_{a_i=1|x_i}\}$, defining the associated Bell operator $\hat{B}_{I^{\langle n \rangle}} = \sum\limits_{x_1x_2...x_n} g_{x_1x_2...x_n} \otimes_{i=1}^{n} A^{(i)}_{x_i}$. For any two pure multipartite qubit states $\ket{\Psi_{\theta_1}^n}$, $\ket{\Psi_{\theta_2}^n}$ with $\theta_1,\theta_2 \in [0,\frac{\pi}{4}]$ \eqref{PureQubits} and $\theta_2 > \theta_1$, if $\Tr (\proj{\Psi_{\theta_1}^n} \hat{B}_{I^{\langle .. \rangle}}) > g_\textnormal{loc}$ then:
\begin{equation}\label{MainResult_ghz}
\Tr (\proj{\Psi_{\theta_2}^n} \hat{B}_{I^{\langle n \rangle}}) > \Tr (\proj{\Psi_{\theta_1}^n} \hat{B}_{I^{\langle n \rangle}}) .
\end{equation}
\end{theorem}
In particular, the theorem implies that if the state $\ket{\Psi_{\theta_1}^n}$ violates the Bell inequality when given measurements are being made on it, the state $\ket{\Psi_{\theta_2}^n}$ does so too with the same measurements.

\begin{proof}
The proof of the above statement follows the structure of the proof of Theorem \ref{THRM:Bell}. By assumption we have that
\begin{align} \label{EQ:ass_multiqubit}
b_{\theta_1}^n \equiv \Tr(\ketbra{\Psi_{\theta_1}^n}{\Psi_{\theta_1}^n} \hat{B}_{I^{\langle n \rangle}}) >g_\textnormal{loc}
\end{align}
As before, we can write, without loss of generality, that 
\begin{equation}\begin{split}\label{EQ:decomposition_multiqubit}
\ket{\Psi_{\theta_1}^n} = \left(
\tfrac{\cos \theta_1 + \sin \theta_1}{\sqrt{2}} \mathbb{1} + \tfrac{\cos \theta_1 - \sin \theta_1}{\sqrt{2}} \sigma_z
\right)\\
\otimes\underbrace{ \mathbb{1} \otimes \ldots \otimes \mathbb{1} }_{n-1} \ket{GHZ^n}
\end{split}\end{equation}
Now, the Bell operator can be decomposed in the Pauli basis as
\begin{equation}\label{MultiDecompB}
\hat{B}_{I^{\langle n \rangle}} = \sum\limits_{\mathbf{i} = 1}^3 c_{i_1...i_n} \sigma_{i_1} \otimes \ldots \otimes \sigma_{i_N}
\end{equation} 
where $\sigma_{i_j}$ denotes one of the Pauli operators $\sigma_x,\sigma_y,\sigma_z$ of $j$-th party. Note that the inequality $I^{\langle n \rangle}$ is a {correlation} inequality, therefore in the above decomposition \eqref{MultiDecompB} none of the operators $\sigma_{i_j}$ can be $\mathbb{1}$. Using this fact and expression \eqref{EQ:decomposition_multiqubit}, the left hand-side of \eqref{EQ:ass_multiqubit} can be written as
\begin{equation}\label{Thrm:mono}
b_{\theta_1}^n = \frac{b_+^n + b_-^n}{2} + \frac{\sin 2\theta_1}{2} (b_+^n - b_-^n) >g_\textnormal{loc},
\end{equation}
where $b_+^n \equiv \bra{GHZ^n}\hat{B}_{I^{\langle n \rangle}}  \ket{GHZ^n}$ denotes the expectation value of $\hat{B}_{I^{\langle n \rangle}}$ on the maximally entangled GHZ state, and similarly $b_-^n \equiv \bra{GHZ_-^n}\hat{B}_{I^{\langle n \rangle}}  \ket{GHZ_-^n}$ for the GHZ state with a relative $-$ sign. Note that this decomposition holds if and only if the number of parties $n$ is even -- as it can be verified that all the cross terms involving $\underbrace{ \mathbb{1} \otimes \sigma_{i_1}\otimes \ldots \otimes \sigma_{i_n}}_{n}$ disappear only when $n$ is even.

{Similarly to the proof of Theorem \ref{THRM:Bell}, observe that the term $\frac{b_+^n + b_-^n}{2}$ from \eqref{Thrm:mono} is the expectation value of 
$\hat{B}_{I^{\langle n \rangle}}$ on a separable state, $\frac{1}{2}(\ketbra{0\dots 0}{0\dots 0} + \ketbra{1\dots 1}{1\dots 1})$, and therefore it is necessarily smaller or equal to $g_\text{loc}$. Since by assumption $b_{\theta_1}^n > g_\text{loc}$, it follows that $b_+^n > b_-^n$ since $\sin(2\theta_1) > 0$ for all $\theta_1 \in [0,\frac{\pi}{4}]$.}

\end{proof}

{Interestingly, in the multipartite scenario the implications of Theorem \ref{THRM:Bell_ghz} become richer than those of Theorem \ref{THRM:Bell} in the bipartite scenario. Specifically, in the multipartite scenario there exist other notions of nonlocality, giving rise to a hierarchy of multipartite correlations as captured by notions such as $k$-producibility or correlation depth \cite{bancal2009quantifying,curchod2015quantifying}. Observe, however, that in the proof of Theorem \ref{THRM:Bell_ghz} our derivation is independent of the type of multipartite nonlocality that is witnessed by the violation of a given correlation Bell inequality. This observation is possible due to the fact that the term $\frac{b^n_+ - b^n_-}{2}$ in \eqref{Thrm:mono} is the expectation value of the inequality on a fully separable state $\frac{1}{2}(\ket{0\dots0}\bra{0\dots0}+\ket{1\dots1}\bra{1\dots1})$. Hence, this term alone can not violate any Bell inequality as the generated correlations are (fully) local. Therefore, our theorem applies to any type of generalized multipartite nonlocality. In particular, if some measurements lead to $k$-partite nonlocal correlations violating a correlation Bell inequality when made on the state $\ket{\Psi^n_{\theta_1}}$, they also generate $k$-partite nonlocal correlations on any state $\ket{\Psi^n_{\theta_2}}$ with $\theta_2 \geq \theta_1$.}

In light of the above theorem, when using {correlation} $n$-partite inequalities to witness nonlocality, for even $n$, the set of measurements leading to nonlocal behaviors when performed on $\ket{\Psi_{\theta_1}^n}$ is included in the set of measurements leading to nonlocal correlations when made on $\ket{\Psi_{\theta_2}^n}$ if $\theta_2 > \theta_1$. In particular, the set of measurements leading to nonlocal correlations on the maximally entangled state $\ket{GHZ^n}$ is the largest
\begin{align}
\mathcal{V}_{\langle n \rangle} (\ket{\Psi_{\theta}^n}) \subseteq \mathcal{V}_{\langle n \rangle} (\ket{GHZ^n}).
\end{align}
where $\mathcal{V}_{\langle n \rangle}$ denotes the set of measurements leading to nonlocal behaviors exhibited with {correlation} inequalities. In the end, the nonlocal volume \eqref{DefAnomaly} is always maximised by the maximally entangled $n$-partite GHZ state  \eqref{GHZ}
\begin{align}
P_{\langle n \rangle} (\ket{\Psi_{\theta}^n}) \leq P_{\langle n \rangle} (\ket{GHZ^n}).
\end{align}
Note that these results are consistent with the numerical findings of \cite{Gdansk}. We leave open the problem of proving Theorem \ref{THRM:Bell_ghz} for an odd number of parties.

\section{CONCLUSIONS}\label{subsec:conclusion}

The nonlocal volume is a measure of nonlocality with a clear operational meaning that seems to establish a one-to-one correspondence between maximal entanglement and maximal quantum nonlocality. Based on the existing results, it is tempting to conjecture that in bipartite systems the maximally entangled state maximizes the nonlocal volume, which would solve the anomaly observed between entanglement and nonlocality when using other measures. In our work, we provide the first analytical results in this direction. Solving the problem in full generality appears challenging because the nonlocal volume is a rather hard function to deal with. Beyond analytical results, it is also worth performing more numerical searches supporting the conjecture, by extending it to more complex scenarios involving more measurements, outputs, or non-projective measurements. The multipartite case is quite unexplored and also contains intriguing questions.

Before concluding, we would like to briefly mention that no anomalies can be seen in the case of steering, where one of the parties has control over the state received and over the measurements performed \cite{Wiseman,SteerQuantif}. In this framework one can see that the set of measurements leading to steering on a partially entangled state is always included in the set of measurements doing the same on the maximally entangled one. This observation holds for any number of measurements, any type of measurements and any dimension $d$. In fact, the probability to violate a steering inequality is always 1 for any pure entangled states, since the set of compatible measurements has mesure zero and therefore random measurements always produce a violation of a steering inequality when performed on any pure entangled state~\cite{DaniPaul:review}.

\section*{Acknowledgements}
We acknowledge fruitful discussions with Michal Oszmaniec. VL is supported by STW Netherlands, and NWO VIDI grant, and an ERC Starting grant. This work is supported by the ERC CoG QITBOX, the AXA Chair in Quantum Information Science, the Mexican Conacyt Graduate Fellowship Program, the Spanish MINECO (QIBEQI FIS2016-80773-P and Severo Ochoa SEV-2015-0522), Fundacio Cellex, Generalitat de Catalunya (CERCA Program and SGR1381).

\section{Appendix}

\subsection{Bell inequalities with single body correlators: violation with measurements on a partially entangled state only}\label{APP:counterexample}

\begin{figure}[!h]
\includegraphics[width=0.5\textwidth]{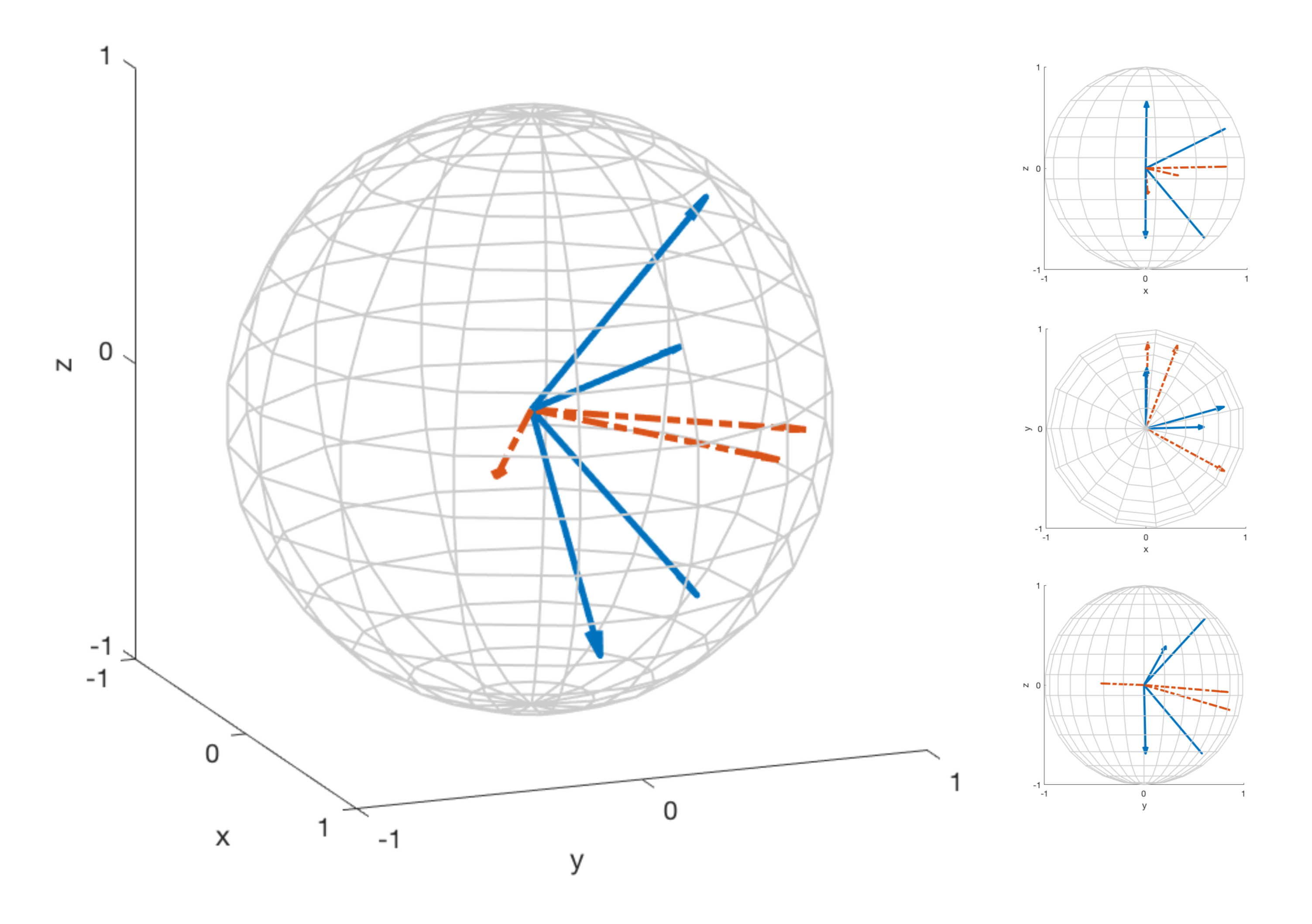}
\caption{Bloch vectors reproducing qubit counterexample measurement settings in the scenario $m_A = 3$, $m_B = 4$ for $\theta = \tfrac{3\pi}{16}$. Red solid vectors correspond to A's settings, blue dashed to B's. On the right projections to xz-plane, xy-plane and yz -plane are presented.}\label{FIG:counter_Bloch}
\end{figure}

Using linear programming, we obtained an example of particular local measurements that do not lead to nonlocality when made on the maximally entangled state, but do so on a partially entangled one. We checked that with our example, the inequality which is violated by the partially entangled state \eqref{PureQubits} with $\theta = \tfrac{3\pi}{16}$ contains single-body correlators, see Table \ref{TAB:counter_ineq}. Table \ref{TAB:counter_AB_vec} presents Bloch vectors corresponding to Alice's and Bob's measurement settings, whereas Figure \ref{FIG:counter_Bloch} visualizes these vectors in the Bloch sphere.

\begin{table}[!h]
\centering
    \caption{The violated Bell inequality in the counterexample case organized in the Collins-Gisin correlator table. Entanglement parameter $\theta = \tfrac{3\pi}{16}$.}
    \begin{tabular}{r|rrr}
    & $\langle A_0 \rangle$ & $\langle A_1 \rangle$ & $\langle A_2 \rangle$ \\ \hline
    $\langle B_0 \rangle$ & $\langle A_0B_0 \rangle$ & $\langle A_1B_0 \rangle$ & $\langle A_2B_0 \rangle$ \\
    $\langle B_1 \rangle$ & $\langle A_0B_1 \rangle$ & $\langle A_1B_1 \rangle$ & $\langle A_2B_1 \rangle$ \\
    $\langle B_ 2\rangle$ & $\langle A_0B_2 \rangle$ & $\langle A_1B_2 \rangle$ & $\langle A_2B_2 \rangle$ \\
    $\langle B_ 3\rangle$ & $\langle A_0B_3 \rangle$ & $\langle A_1B_3 \rangle$ & $\langle A_2B_3 \rangle$ \\
    \end{tabular}\\ \vspace{0.5cm}

    $=$ \vspace{0.3cm}\begin{tabular}{r|rrr}
    & $-0.25$ & $0$ & $0.25$ \\ \hline
    $-0.13$ & $0.25$ & $-0.25$ & $-0.25$ \\
    $-0.13$ & $0.25$ & $0.25$ & $-0.25$ \\
    $-0.01$ & $0$ & $0$ & $0$ \\
    $0$ & $-0.25$ & $0$ & $0.25$ \\
    \end{tabular}
    \label{TAB:counter_ineq}
\end{table}

\begin{table}[!h]
\footnotesize
\caption{Bloch vectors corresponding to measurement settings for A and B leading to the counterexample. Entanglement parameter $\theta = \tfrac{3\pi}{16}$.}
    \label{TAB:counter_AB_vec}
    \begin{minipage}{.48\linewidth}
    \centering
    \caption*{Alice's measurements.}
    \begin{tabular}{r| r r r}
    &$x = 0$ & $x = 1$ & $x = 2$ \\ \hline
    $\sigma_x$& $0.0213$  &  $0.3539$ &   $0.8786$\\
    $\sigma_y$& $0.9599$  &  $0.9320$  & $-0.4772$\\
    $\sigma_z$& $-0.2795$ & $ -0.0780$  &  $0.0176$\\
    \end{tabular}
    \end{minipage}
    \begin{minipage}{.48\linewidth}
    \centering
    \caption*{Bob's measurements.}
    \begin{tabular}{r r r r}
    $y = 0$ & $y = 1$ & $y = 2$ & $y = 3$ \\ \hline
    $0.8685$  &   $0.0095$  & $-0.0025$  &  $0.6437$ \\
    $0.2420$  &  $0.6762$  &  $0.6456$  &  $0.0175$ \\
    $0.4326$  &  $0.7367$  & $-0.7636$ &  $-0.7651$ \\
    \end{tabular}
    \end{minipage}
\end{table}

\end{document}